\documentclass{aa}
\usepackage{graphics}

\begin{document}
\thesaurus{08.05.1, 08.23.2, 
           09.04.1, 09.08.1, 09.09.1, 09.11.1,
           11.09.1, 11.13.1  } 

\def\frac{$''$\hspace*{-.1cm}}
\def\deg{$^{\circ}$}
\def\min{$'$}
\def\deg{$^{\circ}$\hspace*{-.1cm}}
\def\min{$'$\hspace*{-.1cm}}
\def\h2{H\,{\sc ii}}
\def\hi{H\,{\sc i}}
\def\hb{H$\beta$}
\def\ha{H$\alpha$}
\def\sm{$M_{\odot}$}
\def\sl{$L_{\odot}$}
\def\ab{$\sim$}
\def\x{$\times$}
\def\sec{s$^{-1}$}

\title{Very young massive stars in the Small 
   Magellanic Cloud, revealed by {\it HST}\,\thanks{Based on observations with 
   the NASA/ESA Hubble Space Telescope obtained at the Space Telescope 
   Science Institute, which is operated by the Association of Universities 
   for Research in Astronomy, Inc., under NASA contract NAS\,5-26555.}}

\offprints{M. Heydari-Malayeri, heydari@obspm.fr}

\date{Received:  December 1, 1998 / Accepted: January 19, 1999}

\titlerunning{SMC N81}
\authorrunning{Heydari-Malayeri et al.}

\author{M. Heydari-Malayeri\inst{1} \and M.R. Rosa\inst{2,}\,\thanks{
    Affiliated to the Astrophysics Division, Space Science Department of
     the European Space Agency.} \and
   H. Zinnecker\inst{3} \and L. Deharveng\inst{4}  
   \and V. Charmandaris\inst{1}}

\institute{{\sc demirm}, Observatoire de Paris, 61 Avenue de l'Observatoire, 
F-75014 Paris, France \and
Space Telescope European Coordinating Facility, European 
Southern Observatory, Karl-Schwarzschild-Strasse-2, D-85748 Garching bei 
M\"unchen, Germany \and
Astrophysikalisches Institut Potsdam, An der Sternwarte 16, 
D-14482 Potsdam, Germany \and
Observatoire de Marseille, 2 Place Le Verrier, 
F-13248 Marseille Cedex 4, France} 

\maketitle

\begin{abstract}
High spatial resolution imaging with the {\it Hubble Space Telescope}
allowed us to  
resolve the compact \h2\, region N81 lying in  
the Small Magellanic Cloud (SMC). We show the presence of a tight 
cluster of newborn massive stars embedded in this  nebular  
``blob'' of $\sim$\,10\frac\, across.  This is the first time 
the stellar content and internal morphology of such an object  
is uncovered. These are among the
youngest massive stars in this galaxy accessible to direct observations 
at ultraviolet and optical wavelengths. 
Six of them are grouped in the core region of $\sim$\,2\frac\,
diameter, with a pair of the main exciting stars 
in the very center separated by only 0\frac.27 or 0.08 pc.
The images display violent phenomena such as stellar winds, shocks, 
ionization fronts, typical of turbulent starburst regions.
Since the SMC is the most metal-poor 
galaxy observable with very high angular resolution, these 
observations provide important templates for studying star formation in the 
very distant metal-poor galaxies which populate the early Universe.


\end{abstract}

\section{Introduction}

It is now generally believed that massive stars form in dense cores of
molecular clouds. Initially, they are enshrouded in dusty remains of
the molecular material, and, therefore, are not observable in
ultraviolet and visible light. At this stage they can only be detected
indirectly at infrared and radio wavelengths, emitted respectively by
the surrounding dust and by the ionized stellar winds. At a later
stage the far-UV photons dissociate the molecules
and ionize the atoms creating ultracompact \h2 regions. Eventually,
the natal molecular cloud is ionized to become a compact \h2
region. As the ionized volume of gas increases, the advancing
ionization front of the \h2 region reaches the cloud border. The
ionized gas then flows away into the interstellar medium according to
the so-called champagne effect (Tenorio-Tagle  \cite{teno}, 
Bodenheimer et
al. \cite{boden}). From this time on the opacity drops and the newborn
stars become accessible to observation in the ultraviolet and visible. \\

The youngest \h2 regions that can be
penetrated with ultraviolet and optical instruments provide 
therefore the best
opportunities for a {\it direct} access to massive stars at very early
stages of their evolution. Because of the small timescales involved,
it is difficult to catch the most massive stars just at this very
point in their evolution, namely when the young \h2 regions emerge
from the associated molecular clouds. Contrary to the situation in the
Galaxy, where interstellar extinction in the line of sight is
generally high, the Magellanic Clouds, especially the SMC, provide an 
environment where the
sites of massive star formation are accessible without additional
foreground extinction. \\

Our search for such very young, emerging \h2 regions in the Magellanic
Clouds started almost a decade ago on the basis of ground-based
observations at the European Southern Observatory. The result was the
discovery of a distinct and very rare class of \h2 regions in the
Magellanic Clouds, that we called high-excitation compact
\h2 ``blobs'' (HEBs). The reason for this terminology was that 
no features could be distinguished with those telescopes.  So far only
five HEBs have been found in the LMC: N159-5, N160A1, N160A2,
N83B-1, and N11A (Heydari-Malayeri \& Testor 1982, 1983, 1985, 1986,
Heydari-Malayeri et al.\,1990) and two in the SMC: N88A and N81
(Testor \& Pakull 1985, Heydari-Malayeri et al.\,1988a). \\

In contrast to the typical \h2 regions of the Magellanic Clouds, which
are extended structures (sizes of several arc minutes corresponding to
more than 50\,pc, powered by a large number of exciting stars), HEBs
are very dense and small regions ($\sim$\,5\frac\, to 10\frac\, in
diameter corresponding to $\sim$\,1.5--3.0\,pc). 
HEBs are, in general, heavily affected by local dust (Heydari-Malayeri
et al. \cite{hey88a}, Israel \& Koornneef \cite{ik91}).
They are probably the final stages in the evolution of the
ultracompact \h2 regions whose Galactic counterparts are detected only
at infrared and radio frequencies (Churchwell \cite{chur}).
Because of the contamination by strong nebular background no direct
information about the exciting stars of HEBs has been achievable with
ground-based telescopes. Furthermore, it is not known whether a single
hot object or several less  massive stars are at work there.  \\

\begin{figure*}
\begin{center}
\resizebox{17cm}{!}{\includegraphics{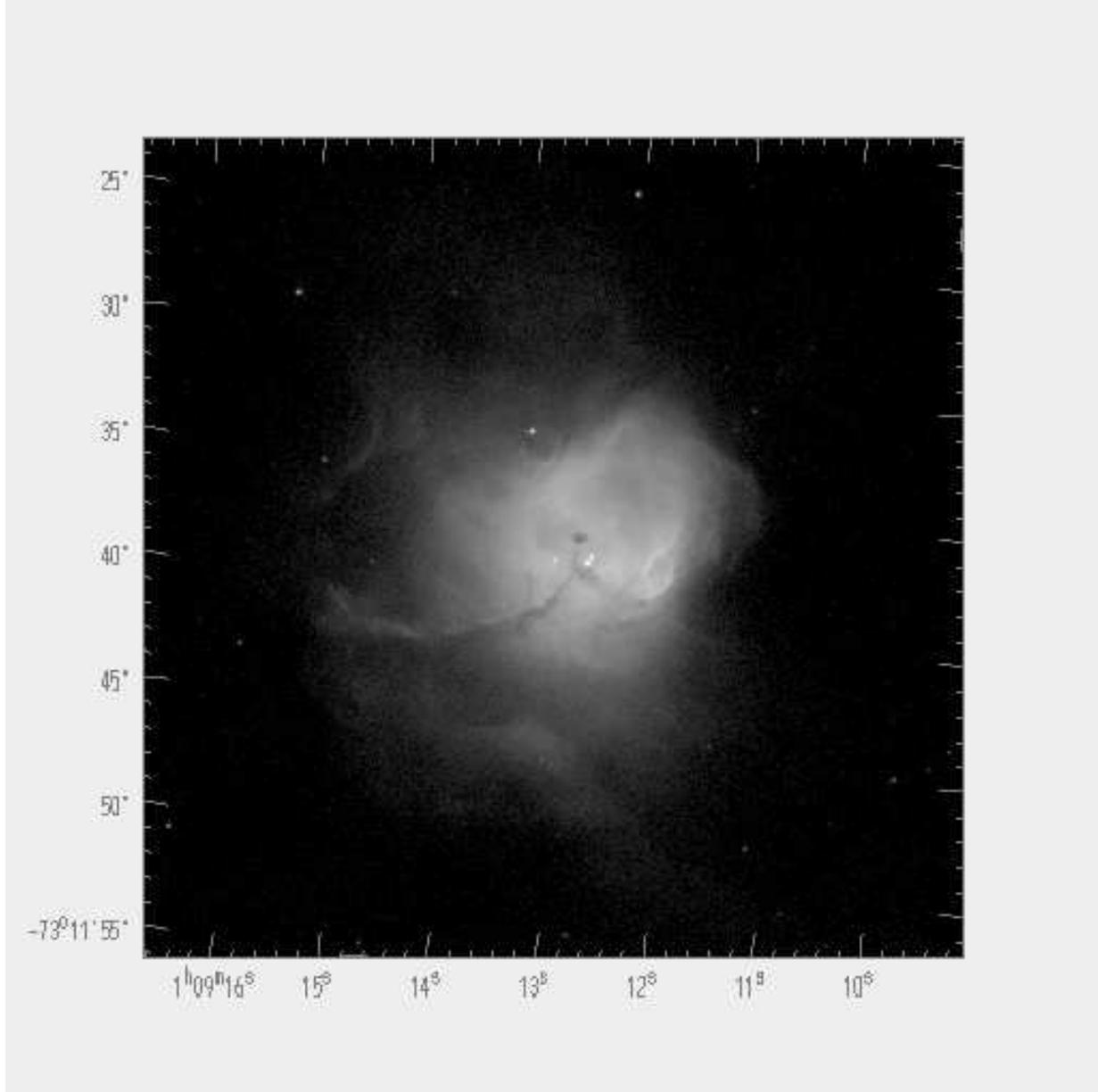}}
\caption{The Small 
Magellanic Cloud compact nebula N81 in \ha\, emission. The two 
bright stars (\#1 \& \#2, see Fig.\,\ref{numbers}) 
lying below the small central absorption ``hole'' are the main 
exciting stars. See Fig.\,\ref{chico} for a close-up of the inner regions. 
The field size is \ab\,33\frac\,$\times$\,33\frac\, 
(\ab\,10\,\x\,10\,pc). 
North is  up and east is left. }
\label{ha}
\end{center}
\end{figure*}

The compact \h2\, ``blob'' 
N81 (Henize \cite{hen}, other designations: DEM138 in Davies et al. 
\cite{dav}, IC\,1644, HD\,7113, etc.) lies in the Shapley Wing 
at \ab\,1\deg.2 (\ab\,1.2\,kpc) from the main body of the SMC. 
Other \h2\, regions lying towards the Wing are from west to 
east N83, N84, N88, N89, and N90. A study of N81 carried out a decade ago 
(Heydari-Malayeri et al.
\cite{hey88a}),  revealed some of its physical characteristics: 
age of 1 to 2.5 million years, mass of ionized gas amounting to 
$\sim$\,350\,\sm, low metal content typical of the chemical 
composition of the SMC, gas density of $\sim$\,500\,cm$^{-3}$, 
electron temperature of 14\,100$^{\circ}$K, etc. 
However, the study suffered from a lack of sufficient 
spatial resolution and could not be pursued by available Earth-bound 
facilities. More specifically, the exciting 
star(s)  remained hidden inside the ionized gas. 
It was not possible to constrain the theoretical models  
as to the nature of the exciting source(s) and choose among various 
alternatives (Heydari-Malayeri et al. \cite{hey88a}). This is, however, 
a critical question for theories of star formation. \\

The use of {\it HST} is therefore essential for advancing our 
knowledge of these objects. 
Here we present the results of our project GO\,6535 
dedicated to direct imaging and photometry of the ``blobs'' as a first 
step in their high-resolution study. A ``true-color'' high-resolution 
image and a brief account of the results for 
the layman were presented in a NASA/HST/ESA Press Release 98-25, 
July 23, 1998 (Heydari-Malayeri et al. \cite{hey98}: 
http://oposite.stsci.edu/pubinfo/pr/1998/25). \\

\section{Observations and data reduction}

The observations of N81 described in this paper were obtained with
the Wide Field Planetary Camera\,2 (WFPC2) on board the {\it HST} on
September 4, 1997.  The small size of N81 makes it an ideal target for
the 36\frac\,-field of WFPC2.  We used several wide- and narrow-band
filters (F300W, F467M, F410M, F547M, F469N, F487N, F502N, F656N,
F814W) with two sets of exposure times (short and long). In each set
we repeated the exposures twice with \ab\,5 pixel offset shifts in both
horizontal and vertical direction. This so-called dithering technique
allowed us to subsequently enhance the sampling of the 
point spread function (PSF) and improve
our spatial resolution by \ab\,20\%.  
The short exposures, aimed at
avoiding saturation of the CCD by the brightest sources, range from
0.6 sec (F547M) to 20 sec (F656N). 
The long exposures  range from 8 sec (F547M) to 300 sec 
(F656N \& F469N).  \\

The data were processed through the standard {\it HST} pipeline
calibration.  Multiple dithered images where co-added
using the {\sc stsdas} task {\it drizzle} (Fruchter \& Hook 
\cite{fru}), while cosmic rays were detected and removed with the {\sc
stsdas} task {\it crrej}.  Normalized images were then created 
using the total exposure times for each filter.
To extract the positions of the stars, the
routine {\it daofind} was applied to the images by setting the
determination threshold to 5$\sigma$ above the local background level.
The photometry was performed setting a circular aperture of 3--4
pixels in radius in the {\it daophot} package in {\sc stsdas}. \\

A crucial point in our data reduction was the sky subtraction. For
most isolated stars the sky level was estimated and subtracted
automatically using an annulus of 6--8 pixel width around each star.
However this could not be done for several stars located in the
central region of N81 due to their proximity. In those cases we
carefully examined the PSF size of each  individual star 
({\sc fwhm}\,\ab\,2 pixels, corresponding to 0\frac.09 on the sky) 
and did an
appropriate sky subtraction using the mean of several nearby off-star
positions.  To convert into a magnitude scale we used zero points in
the Vegamag system, that is the system where Vega is set to zero mag 
in Cousin broad-band filters.  The magnitudes measured were corrected for
geometrical distortion, finite aperture size (Holtzman et al. \cite{holtz}),
and charge transfer efficiency as recommended by the HST Data
Handbook.  The photometric errors estimated by {\it daophot} are
smaller than 0.01 mag for the brighter (14--15 mag) stars, while they
increase to $\sim$\,0.2 mag for 19 mag stars. \\

\begin{figure*}
\begin{center}
\resizebox{12cm}{!}{\includegraphics{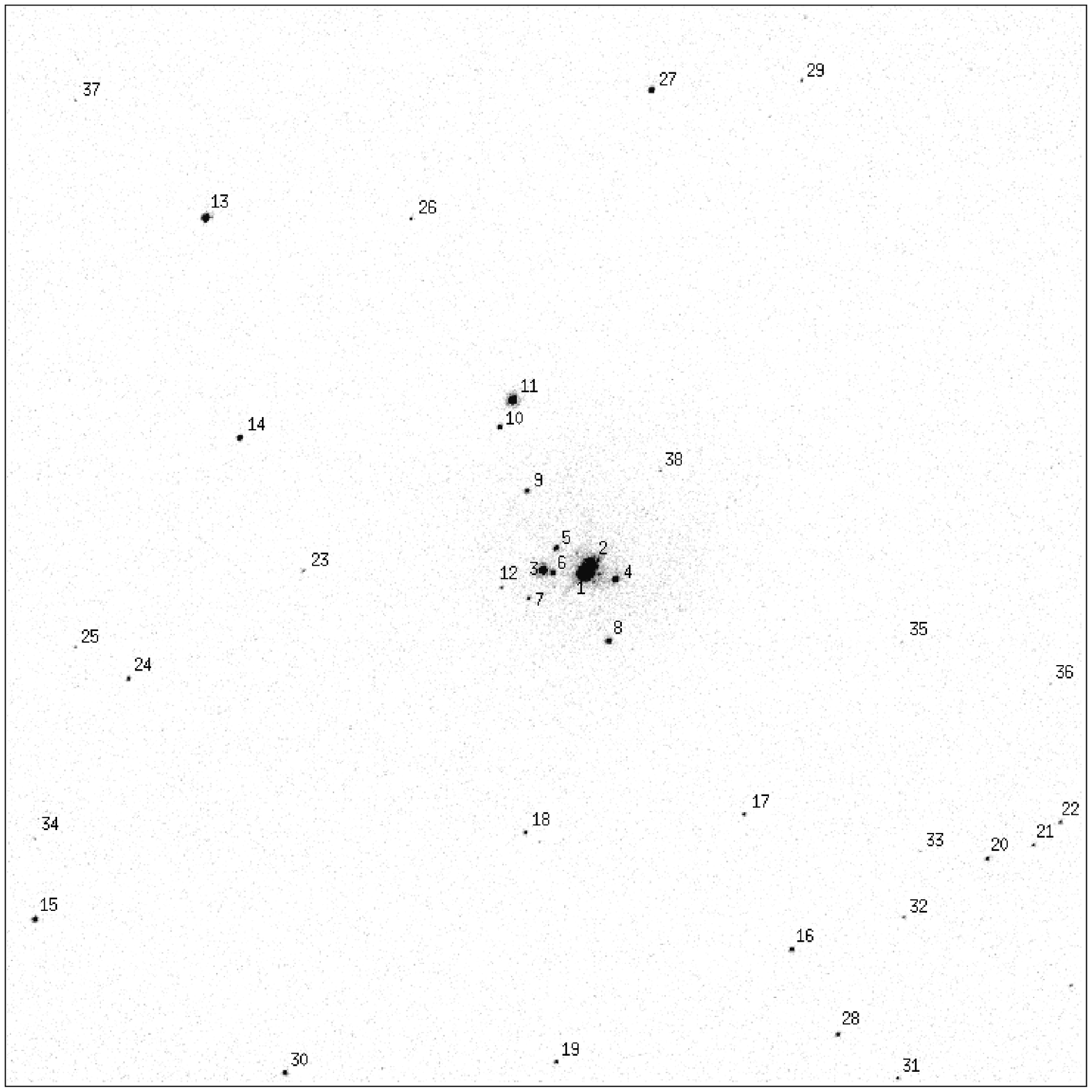}}
\caption{A WFPC2 image taken using  the Str\"omgren $y$ filter where 
only the brighter stars are labelled. The photometry of the 
brightest subsample is presented in Table\,\ref{phot}.   
The field size and orientation are the 
same as in Fig.\,\ref{ha}.}
\label{numbers}
\end{center}
\end{figure*}

We note that the filter F547M is wider than the standard Str\"omgren
$y$ filter. To evaluate the presence of any systematic effects in our
photometry and color magnitude diagrams due to this difference in the
filters, we used the {\sc stsdas} package {\it synphot}.  Using
synthetic spectra of hot stars, with spectral types similar to those
found in \h2\, regions, we estimated the difference due to the {\it
HST} band-passes to be less than 0.002 mag, which is well within the
photometric errors. \\

The ``true-color'' image of N81 (Heydari-Malayeri et al. \cite{hey98}) 
was assembled from three separate WFPC2 images using the {\sc iraf} 
external package {\sc color} task 
{\it rgbsum}. The basic images were the ultraviolet (F300W) and 
the hydrogen emission blue and red lines  \hb\, 
(F487N) and \ha\, (F656N). \\

Two line intensity ratio maps were secured using the normalized 
\ha, \hb\, and [O\,{\sc iii}]\,$\lambda$5007 (F502N) images 
($\S$\,3.2 and $\S$\,3.3). In order to enhance the 
S/N ratio in the fainter parts, each image was first smoothed 
using a 2\,\x\,2-pixel Gaussian filter.  \\

\begin{figure}
\resizebox{\hsize}{!}{\includegraphics{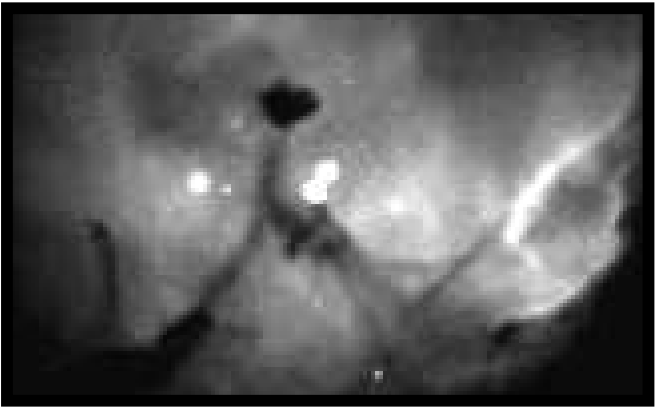}}
\caption{A close-up of the inner part of N81 in \ha\, emission. 
The two brighter stars (\#1 \& \#2), 0\frac.27 apart, and 
the absorption features as well as the western ridges, 
representing ionization/shock fronts, stand out 
prominently.   The inner ridge lies at a projected distance 
of \ab\,0.7\,pc from the two main stars; the outer one 
at \ab\,1.0\,pc.  
Field size \ab\,7\frac.4\,\x\,4\frac.6. (\ab\,2.2\,\x\,1.4\,pc).}
\label{chico}
\end{figure}

\section{Results}

\subsection {Overall view}

The WFPC2 imaging, in particular the $I$ band filter,
reveals some 50 previously unknown stars  lying towards N81, where not 
even one was previously observable. Six of them are grouped in 
the core region of $\sim$\,2\frac\, wide, as displayed in Fig.\,\ref{ha}. 
The brightest ones are identified by numbers in Fig.\,\ref{numbers}.
See also the true-color version of this image accompanying the 
{\it HST} Press Release (Heydari-Malayeri et al. \cite{hey98}).  
Two bright stars (\#1 \& \#2) occupy a central position and  
are probably the main exciting sources of the \h2\, region. 
Only 0\frac.27 apart on the sky (projected separation $\sim$\,0.08\,pc),
they are resolved in the WFPC2 images 
(Fig.\,\ref{chico}).  \\

Two prominent dark lanes divide the nebula into three 
lobes. One of the lanes ends in a magnificent curved plume  
more than 15\frac\, (4.5\,pc) in length. The absorption 
features are probably parts of the molecular cloud 
associated with the \h2\, region ($\S$\,4.2). 
The extinction due to dust grains 
in those directions amounts to \ab\,1 mag as indicated by 
the \ha/\hb\, map ($\S$\,3.2).
A conspicuous absorption ``hole'' or dark globule of radius 
\ab\,0\frac.25 is situated towards the center of the \h2\, region,  
where the extinction reaches even higher values. 
The apparent compact morphology of this globule in the presence of a 
rather violent environment is intriguing. We explore some possibilities
of its origin in $\S$\,4.1.  \\

\begin{figure}
\resizebox{\hsize}{!}{\includegraphics{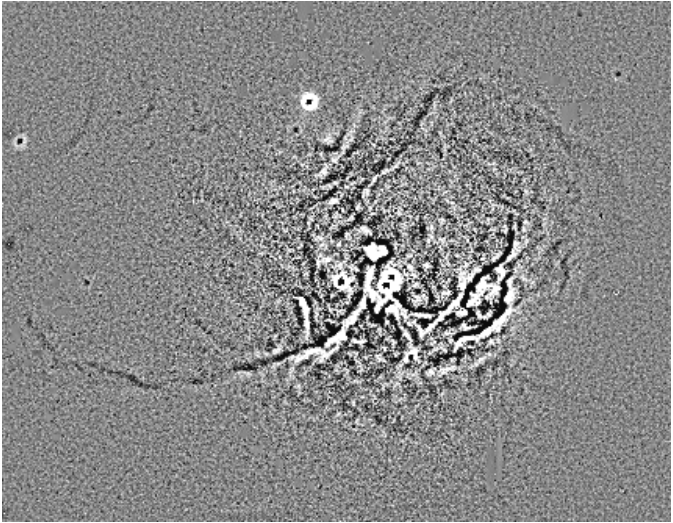}}
\caption{An unsharp masking image of N81 obtained 
in \ha,  which highlights the filamentary  patterns 
of the nebula (see the text). 
Field size \ab\,20\frac\,\x\,16\frac\, (6.0\,\x\,4.7\,pc), orientation 
as in Fig.\,\ref{ha}.}
\label{vent}
\end{figure}

An outstanding aspect is the presence of arched filaments, gaseous 
wisps, and narrow ridges  
produced by powerful stellar winds and 
shocks from the hot, massive stars. 
We are therefore witnessing a very turbulent 
environment typical of young regions of star formation. 
The two bright ridges 
lying west of stars \#1 \& \#2 are probably ionization/shock 
fronts. The filamentary and wind induced structures 
are best seen in Fig.\,\ref{vent}, which presents an unsharp masking 
image of N81 in \ha\, without large scale structures. 
In order to remove these brightness variations and enhance 
the high spatial frequencies, 
a digital ``mask'' was created from the \ha\, image. 
First the \ha\, image was convolved by a 2\,\x\,2-pixel 
Gaussian, and then 
the smoothed frame  was subtracted from the original \ha\, image. 
Interestingly, the inspection of the orientation of these 
arched filaments suggests the presence of at least three sources of stellar 
winds: stars \#1 \& \#2 jointly, \#3, and probably \#11, 
which are the four brightest blue stars of the cluster.  \\

\begin{figure*}
\resizebox{\hsize}{!}{\includegraphics{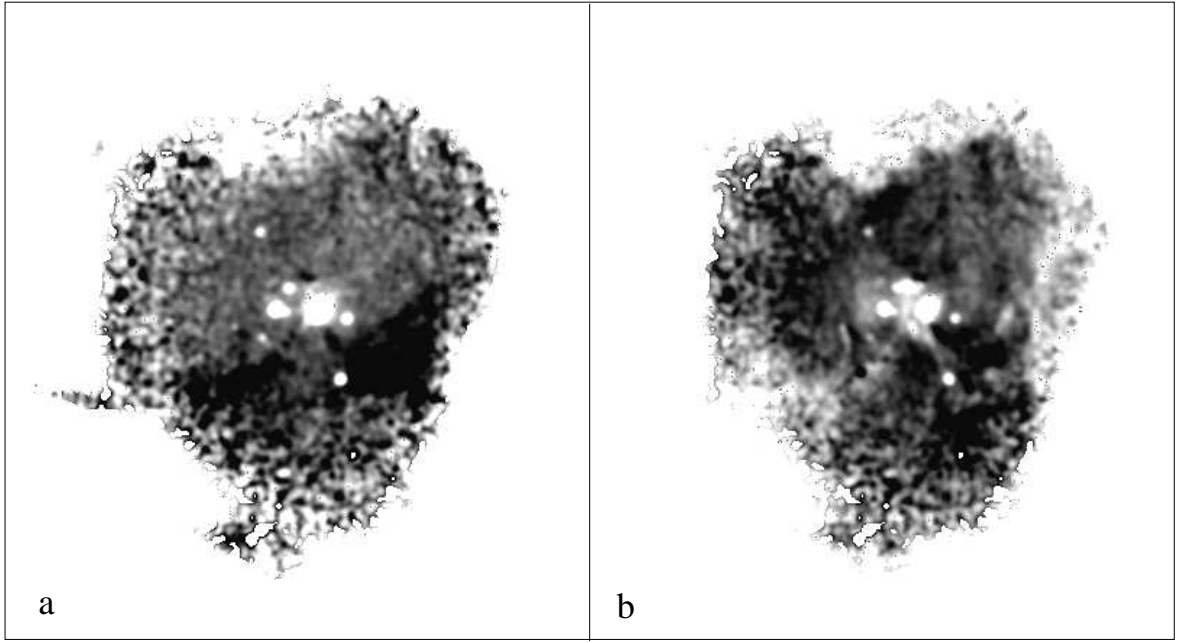}}
\caption{Line intensity ratios in the central part of 
N81. The darker the color the higher the ratio. 
White spots are stars; see Fig.\,\ref{numbers} for their identification. 
{\bf a)} Balmer decrement \ha/\hb. Its mean value over the region 
is 3.30 and the ratio 
goes up to $\sim$\,4.5 towards the central small absorption ``hole''.
The spur extending to the left is the bright end of the 
eastern absorption lane. 
{\bf b)} The [O\,{\sc iii}]$\lambda$5007/\hb\, ratio.   
It is everywhere higher than 4.5 and rises to more than 6. } 
\label{combo}
\end{figure*}

\subsection{Extinction}

A map of the  \ha/\hb\, Balmer decrement is presented 
in Fig.\,\ref{combo}a. The ratio
reaches its highest values towards the dark lanes (\ab\,3.8), the 
western ridges (\ab\,4.5), and  the dark globule 
(\ab\,4.5). This 
latter value is very close to the highest ratio (4.3) 
expected for a medium in which dust is locally mixed with gas.   
The dark ``hole'', while present, does not show up  
prominently in Fig.\,\ref{combo}a,  
because it is small (\ab\,0\frac.25 across corresponding to 
\ab\,15\,000\,AU) and the binning by convolution used to enhance 
the S/N ratio  in the fainter parts ($\S$\,2)
has reduced the line ratio. The high \ha/\hb\, ratios and the fact that 
the interstellar extinction is known to be small towards the SMC
(Pr\'evot et al. \cite{prev}) support the idea that dust is 
local and probably mixed with gas in this young \h2\, region. 
Furthermore, the high 
resolution observations show that the extinction towards 
N81 is generally higher than  previously believed.
A mean \ha/\hb\,=\,3.30   corresponds to $A_{V}$\,=\,0.40 mag  
($c$(\hb)\,=\,0.20), if the interstellar reddening law is used. 
For comparison, the ground-based observations had yielded 
\ha/\hb\, values of 3.05 (Heydari-Malayeri et al. \cite{hey88a}) 
and 2.97 (Caplan et al. \cite{cap}, using a circular diaphragm of 
4\min.89 in diameter). 
We remark that the Balmer  ratio decreases with decreasing 
spatial resolution. This provides another indication 
that dust is concentrated towards the inner parts of N81. \\

\subsection{Nebular emission}

The [O\,{\sc iii}]$\lambda$5007/\hb\, intensity map 
(Fig.\,\ref{combo}b) reveals a relatively extended 
high-excitation zone  with a mean  
value of $\sim$\,4.8.  
The highest intensity ratio, $\sim$\,5.5, 
belongs to the region of shock/ionization fronts represented 
by the two bright western ridges (Fig.\,\ref{chico}). 
It is not excluded that collisional excitation of the O$^{++}$ ions 
by shocks contributes to the high value of the ratio 
in that region. Another high excitation zone runs from the east of 
star \#9 to the south.  
The remarkable extension of the [O\,{\sc iii}]$\lambda$5007/\hb\, ratio 
suggests that the  O$^{++}$ ions  
in N81 occupy almost the same zone as  H$^{+}$.
This is in agreement with our previous chemical abundance 
determination results (Heydari-Malayeri et al. \cite{hey88a}) 
showing that more than 80\% of the total number of oxygen atoms in N81
are in the form of O$^{++}$.  
The extension of the ratio also suggests 
that the \h2\, region is not powered by one central, 
but  by several separate hot stars.  \\

We measure a total \hb\, flux  $F$(\hb)\,=\,7.69\,$\times$\,10$^{-12}$ 
erg cm$^{-2}$ s$^{-1}$ above 3$\sigma$ level 
for N81  without  the stellar contribution 
and accurate to \ab\,3\%.  Correcting for a 
reddening coefficient of $c$(\hb)\,=\,0.20 ($\S$\,3.2) gives 
$F_{0}$(\hb)\,=\,1.20\,$\times$\,10$^{-11}$ erg cm$^{-2}$ s$^{-1}$. 
From this a Lyman continuum flux of 
$N_{L}$\,=\,1.36\,$\times$\,10$^{49}$ photons s$^{-1}$ can be 
worked out if the \h2\, region is assumed to be ionization-bounded. 
A single main sequence star of type O6.5 or O7 can account for 
this ionizing UV flux 
(Vacca et al. \cite{vacca}, Schaerer \& de Koter \cite{sch}). 
However, this is apparently an 
underestimate since the dust grains mixed with gas would 
considerably absorb the UV photons, and moreover the \h2\, 
region is probably 
density-bounded, since one side of it has been torn open towards 
the interstellar medium. \\

\begin{table*}[t]
\caption[]{Photometry of the brightest stars towards N81}
\label{phot}
\begin{flushleft}
\begin{tabular}{ccccccccc}   
\hline
Star number & $\alpha$ & $\delta$ & $v$ & $b$ & $y$ & He\,{\sc ii} & 
   wide $U$ & $I$ \\
            &  (J2000)  & (J2000) & (F410M) & (F467M) & (F547M)  &  
   (F469N)  & (F300W) & (F814W)       \\ 
\hline
1  & 01:09:13.05 & --73:11:38.27 & 14.04 & 14.28 & 14.38 & 14.30 & 12.64 & 15.80\\
2  & 01:09:13.03 & --73:11:38.03 & 14.50 & 14.76 & 14.87 & 14.81 & 12.88 & 14.90\\
3  & 01:09:13.35 & --73:11:38.37 & 15.83 & 16.02 & 16.10 & 16.02 & 14.37 & 16.10\\
4  & 01:09:12.83 & --73:11:38.31 & 17.20 & 17.48 & 17.41 & 17.46 & 15.65 & 17.17\\
5  & 01:09:13.28 & --73:11:37.63 & 18.09 & 18.24 & 18.29 & 18.23 & 16.85 & 18.37\\
6  & 01:09:13.28 & --73:11:38.39 & 18.12 & 18.19 & 18.11 & 18.13 & 17.30 & 17.59\\
7  & 01:09:13.43 & --73:11:39.29 & 19.25 & 19.69 & 19.64 & 19.27 & 18.15 & 19.06\\
8  & 01:09:12.82 & --73:11:40.23 & 18.00 & 17.99 & 17.84 & 17.90 & 16.68 & 17.57\\
9  & 01:09:13.54 & --73:11:36.01 & 18.66 & 18.87 & 18.80 & 18.83 & 17.60 & 18.69\\
10 & 01:09:13.80 & --73:11:34.17 & 18.13 & 18.46 & 18.47 & 18.38 & 17.01 & 18.49\\
11 & 01:09:13.74 & --73:11:33.31 & 15.38 & 15.64 & 15.74 & 15.64 & 13.85 & 16.11\\
13 & 01:09:16.08 & --73:11:29.07 & 16.42 & 16.57 & 16.65 & 16.52 & 15.36 & 16.67\\
14 & 01:09:15.61 & --73:11:35.63 & 17.76 & 17.90 & 17.99 & 17.86 & 16.50 & 18.14\\
15 & 01:09:16.58 & --73:11:51.21 & 18.73 & 18.49 & 18.10 & 18.21 & 19.22 & 17.21\\
16 & 01:09:11.22 & --73:11:48.87 & 18.37 & 18.59 & 18.60 & 18.58 & 17.32 & 18.57\\
17 & 01:09:11.69 & --73:11:45.00 & 19.18 & 19.13 & 19.39 & 19.23 & 18.05 & 19.07\\
18 & 01:09:13.22 & --73:11:46.43 & 19.07 & 19.26 & 19.39 & 19.16 & 18.00 & 19.13\\
24 & 01:09:16.16 & --73:11:43.49 & 20.44 & 19.67 & 19.10 & 20.58 &  --   & 18.10\\
27 & 01:09:13.07 & --73:11:23.24 & 18.81 & 18.38 & 17.74 & 18.13 &  --   & 16.82\\
\hline
\end{tabular}
\end{flushleft}
\end{table*}

\subsection{Stellar content}

\begin{figure*}
\begin{center}
\resizebox{18cm}{!}{\includegraphics{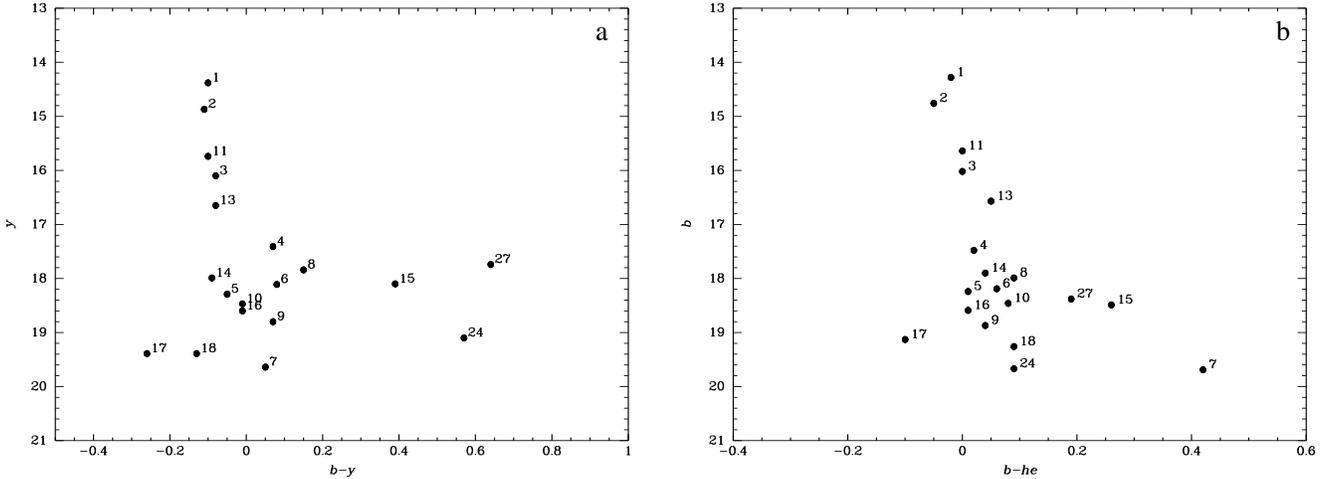}}
\caption{Color-magnitude diagrams of 
the brightest stars revealed towards the SMC compact \h2\, region N81.
{\bf a}) $y$ versus $b$\,--\,$y$, 
based on WFPC2 imaging with the Str\"omgren filters $b$
(F467M) and $y$ (F547).
{\bf b}) $b$ versus $b$\,--\,He\,{\sc ii} 
showing the He\,{\sc ii} 4686 excess, 
based on narrow-band filter F469N and broad-band continuum F467M.} 
\label{cm}
\end{center}
\end{figure*}

The results of the photometry for the brightest stars are presented 
in Table\,\ref{phot}, 
where the star number refers to Fig.\,\ref{numbers}.
The color-magnitude diagram $y$ versus $b - y$  
is displayed in Fig.\,\ref{cm}a.  
It shows a blue cluster centered on Str\"omgren colors 
$b - y = -0.05$, or $v - b = -0.20$, 
typical of massive OB stars 
(Relyea \& Kurucz \cite{rk}, Conti et al. \cite{conti}). 
This is confirmed by the $U$\,--\,$I$ colors deduced from 
Table\,\ref{phot}.   
Almost all these stars lie within the \h2\, region and we 
are in fact viewing a  starburst in this part of the SMC.    
We may neglect some of the ionizing stars if they lie deeper 
in the molecular cloud and are affected by larger extinctions.
The three red stars (\#15, \#24, and \#27), located outside  
the \h2\, region, show up particularly in the true-color  
image (Heydari-Malayeri et al. \cite{hey98}) 
and  are probably evolved stars not belonging 
to the cluster. The two main exciting stars (\#1 and \#2) stand 
out prominently on the top of the color-magnitude diagram. \\

One can estimate the luminosity of the brightest star of 
the cluster (\#1), although in the absence of spectroscopic data  
this is not straightforward. Using a mean reddening of 
$A_{V}$\,=\,0.4 mag ($\S$\,3.2), and a distance modulus 
$M$\,--\,$m$\,=\,19.0 (corresponding to a distance of 63.2\,kpc, e.g. 
Di Benedetto \cite{di} and references therein), 
we find a visual absolute magnitude $M_{V}$\,=\,--5.02 for star \#1. 
If a main sequence star, this corresponds to an O6.5V  according to 
the  calibration of Vacca et al. (\cite{vacca}) for Galactic 
stars. The corresponding luminosity and mass would be 
{\it log L}\,=\,5.49\,\sl\,  and $M$\,=\,41\,\sm.  
The star may be more massive than this, 
since sub-luminosity and/or peculiar 
extinction would be consistent with extreme youth (Walborn et al.
\cite{wal99}). \\

The images obtained through a narrow-band filter (F469N) centered 
on the He\,{\sc ii} 4686\,\AA\, line were compared with those using the 
broad-band filter representing the Str\"omgren\,$b$ (F467M).
The resulting photometry is displayed in the  
color-magnitude diagram shown in Fig.\,\ref{cm}b. 
Interestingly, three stars show an apparent He\,{\sc ii} excess. 
However, one should  note that since two of 
these (\#27 and \#15) are red types (Fig.\,\ref{numbers}, and the 
true-color image in Heydari-Malayeri et al. \cite{hey98}), such a
narrow-band enhancement could also be due to an artifact of molecular
absorption bands throughout the spectra.
On the other hand, the third star (\#7) is blue and 
has stronger apparent He\,{\sc ii} emission. This 
star lies deep in the core of the star cluster/\h2\, region where the 
nebular excitation is high. It  may  be a Wolf-Rayet 
or Ofpe/WN candidate in the SMC. \\

\section{Discussion and concluding remarks}

\subsection{Morphology of N81}

N81 is a young \h2\, region whose moving ionization front 
has reached the surface of its associated molecular cloud 
(see $\S$ 4.2), as predicted by the champagne model 
(Tenorio-Tagle  \cite{teno}, Bodenheimer et al. \cite{boden}).
Ionized gas is pouring out into the interstellar medium 
with a relative  velocity of 4\,\,km\,\sec\, ($\S$4.2).
The bright central core of N81 
is presumably a cavity created by the stellar photons on 
the surface of the molecular cloud. The two bright ridges 
lying at  projected distances of \ab\,0.7 and \ab\,1.0\,pc 
west of stars \#1 \& \#2 (Fig.\,\ref{chico}) 
are probably parts of the cavity seen edge-on. They represent 
ionization/shock fronts advancing in the molecular cloud.  
The higher excitation zones, indicated by 
the [O\,{\sc iii}]/\hb\, map ($\S$\,3.3),  may be 
parts of the cavity surface situated perpendicularly to the line 
of sight. The \h2\, region is probably ionization-bounded in those 
directions. The outer, diffuse areas with fainter brightness 
(Fig.\,\ref{ha})  are the champagne flow in which strong  
winds of massive stars have given rise to the filamentary pattern.\\

The absorption lanes and the 
the dark globule may  represent the denser, optically thick 
remains of the natal molecular cloud which have so far survived the 
action of harsh ultraviolet photons of the exciting stars. 
High resolution imaging observations by 
{\it HST} have shown the presence of 
massive dust pillars inside the Galactic \h2\, region 
M16 (Hester et al. \cite{hest}) and more recently in the LMC 
giant \h2\, region 30 Dor 
(Scowen et al. \cite{sco}, Walborn et al. \cite{wal99}). 
By comparison, the dark globule in N81 may be the summit of such 
a dust pillar in which second generation stars may be forming.
To further investigate this idea, a follow-up high-resolution near-IR 
imaging of the central region is essential. At longer wavelengths 
we are less affected by the absorption and we will be able to probe
deeper into the core of the globule.   \\

\subsection{Molecular cloud}

The molecular cloud associated with N81 has been observed during the 
ESO-SEST survey of the CO emission in the Magellanic Clouds.  
Israel et al. (\cite{is})  detected $^{12}$CO\,(1\.-\.0) emission 
at two points towards N81 using a resolution of 43\frac\, 
(\ab\,13\,pc, or \ab\,4 times the size of the \h2\, region). 
The brighter component has a main beam brightness temperature of 
375\,mK, a line width of 2.6\,km\,\sec\, and a LSR velocity of 
152 km\,\sec. The molecular emission velocity is 
in agreement with the velocity of the ionized gas $V_{LSR}$\,=\,147.8 
km\,\sec\, which we measured  on the basis of 
high dispersion \hb\, spectroscopy  (Heydari-Malayeri et al. 
\cite{hey88a}). The difference of 4.2 km\,\sec\, is probably due   
to the local motion of the ionized gas streaming into the 
interstellar medium towards the observer. The molecular 
cloud is  brighter than those detected towards the 
neighboring \h2\, regions N76, N78, and N80, but is weaker than 
that associated with N84 which has a distinct velocity 
(168 km\,\sec). We do not know the size, the 
morphology, nor the mass of the molecular cloud,  and we cannot
localize  the \h2\, region N81 with respect to it. \\

The SMC is known to have an overall  complex structure with several 
overlapping neutral hydrogen layers (McGee \& Newton \cite{McGee}). 
We used the recent observations by Stanimirovic et al. (\cite{stan})
to examine the \hi\, emission towards N81. The authors 
combine new Parkes telescope 
observations with an ATCA (Australia Telescope Compact Array) aperture 
synthesis mosaic to obtain a set of images sensitive to all angular 
(spatial) scales between 98\frac\, (30\,pc) and 4\deg\, (4\,kpc)
in order to study the large-scale \hi\, structure of the SMC. 
An  \hi\, concentration 
focused on N84/N83 extends eastward until N81 (\ab\,30\min\, apart). 
The \hi\, spectra towards  N81 and N83/N84 
show complex profiles ranging from 
\ab\,100 to 200 km\,\sec\, with two main peaks
at  \ab\,120 and 150 km\,\sec\, but spread over tens of km\,\sec.
The corresponding column densities are 
3.94\,\x\,10$^{21}$ and 5.29\,\x\,10$^{21}$ atoms\, cm$^{-2}$ 
respectively. In consequence, the correlation between the 
\hi\, and CO clouds towards N81 does not seem simple. \\

The WFPC2 images reveal a turbulent nebula in which 
arc-shaped features are sculpted in the ionized 
gas under the action of violent shocks, ionization 
fronts, and stellar winds.  The presence of shocks in 
N81 was evoked by Israel \& Koornneef (\cite{ik88})
in order to explain the infrared molecular hydrogen emission  
which they detected towards this object.  H$_{2}$ emission may 
be caused either by shock excitation due to stars embedded in a 
molecular cloud or by fluorescence of molecular material in the 
ultraviolet radiation field of the OB stars exciting the \h2\, 
region. According to these authors,  shock excitation of H$_{2}$ is 
only expected very close to (within 0.15\,pc of) the stars, 
while radiative excitation can occur at larger distances 
(\ab\,1 to 2\,pc). Our {\it HST} observations suggest that  
the radiative mechanism is the dominant one, since the 
shock/ionization fronts are  situated at projected distances larger 
than 0.15\,pc  from the exciting stars ($\S$\,4.1). \\

\subsection{Star formation}

These observations are a breakthrough in the investigation of 
the exciting source of N81. This  
compact \h2\, region is not powered by a single star, 
but by a small group of newborn massive stars. This result is important 
not only for studying the energy balance of the \h2\, region, but 
also because it presents new evidence in support of 
collective formation of massive stars. 
Recent findings suggest that massive star formation is  probably a collective
process (see, e.g., Larson \cite{lar} and references therein, 
Bonnell et al. \cite{bonn}). This is also in line with the results 
of high-resolution, both ground-based and space observations 
(Weigelt \& Baier \cite{wei},
Heydari-Malayeri et al. \cite{hey88b}, Walborn et
al. \cite{wala}a), in particular the resolution of the so-called
Magellanic supermassive stars into tight clusters of very massive
components (Heydari-Malayeri \& Beuzit \cite{hey94} and references therein). 
It should however be emphasized
that these cases pertain to relatively evolved stellar
clusters. They are not associated with compact \h2 regions, probably
because the hot stars have had enough time to disrupt the gas.  \\

N81 is a  rare case in the SMC since a small cluster of massive
stars is caught almost at birth. It provides a very good chance to 
check the history of massive star evolution (de Koter et al. 
\cite{koter}). Massive stars are believed to enter the main
sequence while still veiled in their natal molecular clouds 
(Yorke \& Kr\"ugel \cite{yk}, Shu et al. \cite{shu},  
Palla \& Stahler \cite{pal}, Beech \& Mitalas \cite{bee}, 
Bernasconi \& Maeder \cite{bern}) implying that these stars may
already experience significant mass loss through a stellar wind, while
still accreting mass from the parental cloud. This point constitutes
an important drawback for current models of massive star evolution
since, contrary to the assumption of the earlier models, a proper
zero-age-main-sequence mass may not exist for these stars (Bernasconi
\& Maeder \cite{bern}). 
As shown by Fig.\,\ref{vent}, the most massive stars of the cluster, 
i.e. stars \#1, \#2, \#11, and \#3,   
seem to be at the origin of stellar winds carving the 
surrounding interstellar medium. Also, if still younger massive 
stars are hidden inside the cental dark globule, they can participate in 
the emission of strong winds.  \\

Another interesting aspect is the small size of the starburst 
that occurred in N81. Apparently, only a dozen massive stars have formed 
during the burst, in contrast to the neighboring region N83/N84 
which is larger and richer (Hill et al. \cite{hill}). The contrast to 
the SMC giant region N66 (NGC346), which has produced a plethora  
of O stars (Massey et al. \cite{mas89}), is even more striking. 
The difference may not be only due to the sizes of 
the original molecular clouds but also to their environments. 
N81 is an isolated object far away from the more active, 
brighter concentrations of matter. Star formation  may be 
a local event  there, while for N66 and its neighboring 
regions, N76, N78, and N80, external factors 
may have played an active role in triggering the starburst.
Judging from their radial velocities (Israel et al. \cite{is}), 
the molecular clouds associated with the \h2\, regions of the Wing 
seem to be independent from each other, and star 
formation has not probably propagated from one side to the other.  
On the other hand, according to Hunter (\cite{hun}),   
massive stars formed in very small star-forming 
regions appear to have a very different mass function, implying 
that different sizes of star-forming events can have 
different massive star products. However, 
the resolution of this question needs more observational data. \\

We may have overlooked a distinct co-spatial population of lower mass stars 
towards N81, as our short exposure WFPC2 images 
were aimed at uncovering the brightest massive stars lying inside
the ionized \h2\, region. Note that the Orion Nebula, which contains 
a low-mass population 
(Herbig \& Terndrup \cite{her}, McCaughrean \& Stauffer 
\cite{McC}, Hillenbrand \cite{hil}),  
would have the same size as N81 if placed in the SMC.
Interestingly, the $I$ band image shows many red stars not 
visible in the blue bands, a few fainter ones 
lying close to stars \#1 and \#2. Do they belong to N81? The answer 
to this question which is crucial for star formation theories 
(Zinnecker et al. \cite{zin}), is not straightforward, 
because of the complex structure of the SMC with its  
overlapping layers (McGee \& Newton \cite{McGee}, Stanimirovic et al. 
\cite{stan}). \\

\subsection{Wolf-Rayet candidate}

Wolf-Rayet stars as
products of massive star evolution are generally very scarce,
particularly so in the metal poor SMC galaxy, which contains only nine
confirmed stars of this category (Morgan et al. \cite{mor}). It is
therefore highly desirable to identify and study every single new
candidate, such as \#7.\\

A noteworthy feature of our new candidate is its apparent
faintness. With $V$\,=\,19.64, it is $\sim$\,3 mag weaker than the
faintest W-R stars in that galaxy detected so far from the ground
(Azzopardi \& Breysacher \cite{azzo}, Morgan et al. \cite{mor}). 
It is also much fainter than the known Of stars in the SMC 
(Walborn et al. \cite{walb}b). Noteworthy as
well is the fact that the small W-R population in the SMC is very
peculiar compared to that in our Galaxy. For example, all nine
W-R stars are binary systems and all, but one, belong to the nitrogen
class. Our candidate may represent the first {\it single} W-R detected in
the SMC. The true nature of this object can only be clarified  with
STIS spectroscopy during the second phase of our project. 
Its confirmation would provide new data for improving massive
star evolutionary models in the low metallicity domain.  \\

\subsection{Future work}

We have presented our first results on the SMC ``blob'' N81 based 
uniquely on direct imaging with WFPC2. The high-resolution 
observations have enabled us to  identify the 
hot star candidates. Forthcoming STIS
observations of these stars will provide the stellar spectra. 
The analysis of the line profiles with non-LTE 
wind models (Schaerer \& de Koter \cite{sch}) will allow us to determine 
the wind properties (terminal velocity, mass loss rate), the 
effective temperature and luminosity, and to derive constraints 
on the surface abundances of H, He, and metals (de Koter et al. 
\cite{koter}, Haser et al. \cite{haser}). The H-R diagrams  so 
constructed will be compared with appropriate low-metallicity 
models (Meynet et al. \cite{mey1}, Meynet \& Maeder \cite{mey2}) 
to yield the evolutionary status of the stars and to subject the 
evolutionary scenarios to observational constraints. \\

Since the SMC is
the most metal-poor galaxy observable with very high angular
resolution, N81 provides an important template
for studying star formation in the very distant metal-poor galaxies
which populate the early Universe.
Although other metal-poor galaxies can be observed with {\it
HST} and their stellar content analyzed from color-magnitude diagrams
(e.g.,\,I\,Zw\,18: 
Hunter \& Thronson \cite{ht}, de Mello et al. \cite{mello}),
the SMC is the most metal-poor galaxy where spectroscopy of individual
stars (required to determine the parameters of massive stars) can be
achieved with the highest spatial resolution. \\

\begin{acknowledgements}
We are grateful to Dr. James Lequeux and Dr. Daniel Schaerer 
who read the manuscript and made insightful comments. 
We would like also to thank an anonymous referee for 
suggestions which helped improve the paper. 
VC would like to acknowledge the financial support from a Marie Curie
fellowship (TMR grant ERBFMBICT960967).
\end{acknowledgements}

{}

\end{document}